\documentclass[aps,prx, amsmath, amssymb, superscriptaddress, twocolumn,longbibliography]{revtex4-1}
\usepackage[left=1in, right=1in, top=1in, bottom=1in]{geometry}
\usepackage[toc,page]{appendix}

\usepackage[latin9]{inputenc}
\usepackage{bm}
\usepackage{tikz,pgfplots}
\usepackage{standalone}
\usepackage{braket}
\usepackage[retainorgcmds]{IEEEtrantools}
\usepackage{graphicx}
\usepackage{mathrsfs}
\usepackage{amsmath}
\usepackage{amssymb}
\usepackage{color}
\usepackage{amsfonts}
\usepackage{times,txfonts}
\usepackage{nicefrac}
\usepackage[colorlinks=true,linkcolor=blue,urlcolor=blue,citecolor=blue,pdfusetitle]{hyperref}
\usepackage{hyperref}
\usepackage[normalem]{ulem}
\usepackage{calrsfs}
\DeclareMathAlphabet{\mathcal}{OMS}{cmsy}{m}{n}
\usepackage{amssymb}
\usepackage{natbib}

\DeclareMathOperator{\tr}{tr}

\begin{document}
	
	\title{Modelling mechanical equilibration processes of closed quantum systems: a case-study}
	
	\author{Sofia Sgroi}
\affiliation{	Centre for Theoretical Atomic, Molecular and Optical Physics, School of Mathematics and Physics, Queen's University Belfast, Belfast BT7 1NN, United Kingdom}
			\author{Mauro Paternostro}
\affiliation{	Centre for Theoretical Atomic, Molecular and Optical Physics, School of Mathematics and Physics, Queen's University Belfast, Belfast BT7 1NN, United Kingdom}

	\date{\today{}}

\begin{abstract}
We model the dynamics of a closed quantum system brought out of mechanical equilibrium, undergoing a non-driven, spontaneous, thermodynamic transformation. In particular, we consider a quantum particle in a box with a moving and insulating wall, subjected to a constant external pressure. Under the assumption that the wall undergoes classical dynamics, we obtain a system of differential equations that describes the evolution of the quantum system and the motion of the wall. We study the dynamics of such system and the thermodynamics of the process of compression and expansion of the box. Our approach is able to capture several properties of the thermodynamic transformations considered and goes beyond a description in terms of an ad-hoc time-dependent Hamiltonian, considering instead the mutual interactions between the dynamics of the quantum system and the parameters of its Hamiltonian.
\end{abstract}

\maketitle{}
	Finite-time work extraction protocols from a system can be realized through external non quasi-static changes of some system parameters \cite{PhysRevLett.78.2690}. Description of work in terms of time-dependent changes of the system Hamiltonian allows us to formulate a quantum version of the first law of thermodynamics \cite{NatCom2018, VinjanampathyAnders, DeffnerCampbell}. Such description is also well suited for the design and optimization of thermal engines when system parameters can be driven during some branches of the thermodynamic cycle \cite{VinjanampathyAnders, DeffnerCampbell, Barontini2019, SciRep2014}. In contrast, heat exchange is undestood in terms of interactions between the system and a reservoir, captured by open quantum system dynamics \cite{Marksman}, which does not induce time-dependent changes in the system Hamiltonian.

	However, any thermodynamic transformation is ultimately due to interactions between two or more systems and most of these transformations occur spontaneously in nature. Driven changes in the parameters of a system Hamiltonian and isochoric heat exchange describe only a subset of possible transformations: in particular, a time dependent Hamiltonian, alone, is not able to describe non-driven transformations which involve equilibratium of systems initially out of mechanical equilibrium unless we know in advance how its time-dependent parameters evolve.

	In this work we present a case-study of a quantum system undergoing such a spontaneous transformation. We start from the assumption that the system dynamics can still be described by time-dependent changes of some parameters of the Hamiltonian, disregarding their possible fluctuations, and hence we assume that the quantum system is interacting with a classical system and that the interaction is mediated by another classical system.
	In particular, we consider a quantum particle in a box with a moving insulating wall. The wall undergoes a classical dynamics and is subjected to internal pressure from the quantum particle, external constant pressure and viscous friction. This configuration can be considered a crude approximation of a particle in a box interacting with a classical gas on the outside, and reminds of the typical thermodynamic problem of a gas inside a piston subjected to a constant external pressure. Despite its simplicity, our model successfully addresses an essential yet often overlooked problem in the general formulation of the framework for the thermodynamics of quantum systems, and allows us to make new considerations on the dynamics of out-of-equilibrium quantum systems.

	In Sec.~\ref{Physical system} we describe the physical system in details and we derive the equation of motions. Numerical simulations of the system dynamics have been carried out and are reported in Sec.~\ref{Numerical simulations} along with examinations of the results. In Sec.~\ref{Considerations an limitations} we discuss our work assumptions, novelties and limitations.

	\section{Physical system}\label{Physical system}
	We consider a particle of mass $m$ in a box of initial length $L_0$. We assume the length of the box at each time $t$, $L(t)$, to be a classical variable. One of the two wall is fixed while the other is free to move. Let us denote the mass of the moving wall with M and its section with $\Sigma$. The particle exerts a force $\Sigma P$ on one side of the wall, while a classical gas at a pressure $P_0$ acts on the other side. Additionally, when the wall is expanding, a classical viscous frictional force $-\gamma \dot L$ causes the wall to decelerate.
	
	Hence, the equation of motion for the wall can be written as
	\begin{equation}\label{wall_Newton}
	m\,{\ddot L}=-\gamma h({\dot L})\,{\dot L}+\Sigma(P-P_0),
%	m\frac{d^2L}{dt^2}=\Sigma(P-P_0)-\gamma \frac{dL}{dt}h\bigg(\frac{dL}{dt}\bigg),
	\end{equation}
	where $h(x)$ is the Heaviside step function.
	We assume the pressure exerted by the quantum particle on the wall to be (cf. Appendix \ref{Appendix1})
	\begin{equation}
	P=\frac{\langle \hat{p}^2\rangle}{m\,L\,\Sigma},
	\end{equation}
	where $\langle \hat{p}^2\rangle$ is the expectation value of the observable $\hat{p}^2$.
	Notice that Eq.~\eqref{wall_Newton} can also be written as
	\begin{equation}\label{wall_Newton_Hamiltonian}
	m\,\ddot{L}=2U/{L}-\Sigma P_0-\gamma\, h\,(\dot{L})\,\dot L,
	\end{equation}
	where $U=\langle\hat{H}\rangle$ is the internal energy of the particle, which is identified with the expectation value of the Hamiltonian  $\hat H$ of the particle~\cite{VinjanampathyAnders}. In the stationary limit ($t\rightarrow+\infty$, $\dot{L}\to0$, $\ddot{L}\to0$), we have %the last equation becomes 
	$L\,\Sigma\, P_0 = PV = 2U$.
	In the classical limit, using the equipartition principle for a classical gas in one dimension, we have $U = {k_BT}/{2}$, so that we obtain the expected equation of state $PV = k_BT$.
	
	In order to describe the evolution of the system we can combine Eq.~\eqref{wall_Newton} with the equation of motion of the quantum particle, which depends on the classical variable $L(t)$ and, at the same time, determines the quantum state of the system, and thus the value of $\langle \hat{p}^2\rangle$.
	
	The evolution of the wavefunction $\psi(x,t)$ of the particle is described using the time-dependent Schr\"{o}dinger equation
	\begin{equation}\label{particle_Schrodinger}
	i\hbar{\partial_t\psi(x,t)}=\frac{\hat{p}^2}{2m}\psi(x,t)
	\end{equation}
	for $x\in[0, L(t)]$ with the boundary condition $\psi(x,t)=0$ elsewhere. %, where $\psi(x,t)$ is the wavefunction of the particle.
	Notice that Eq.~\eqref{particle_Schrodinger} has to be taken with some caution, as the domain of the wavefunction is changing with time and the derivative $\partial_t\psi(x,t)$ has to be defined properly (we cannot define a sum of vectors belonging to two different Hilbert spaces).
	
	A possible strategy to overcome this problem is to find a unitary transformation that allows us to describe the evolution using a common fixed domain for the wavefunction at different times. In our case, this can be achieved considering a dilation $x\rightarrow x/L(t)$, implemented by the unitary transformation $U(L)\,\psi(z)=\sqrt{L}\,\psi(Lz)$. This approach leads us to a new equation of motion with fixed boundary conditions and where $L(t)$ is a simple time-dependent parameter. Such equation reads~\cite{Martino2013}
	\begin{equation}\label{fixed_bound}
	i\hbar\partial_t\phi(x,t)=\left[\frac{\hat{p}^2}{2mL^2(t)}-\frac{\dot{L}(t)}{2L(t)}(\hat{x}\hat{p}+\hat{p}\hat{x})\right]\phi(x,t)
	\end{equation}
	for $x\in[0,1]$, where $\phi(x,t)$ is the wavefunction of the particle in this new representation, and we have taken $L(0)=1$. We can derive an equivalent equation in terms of the eigenbasis $\{|n\rangle\}$ of the initial Hamiltonian $\hat{H}(0)$ upon which we decompose the wavefunction as $\phi(x,t)=\sum_n c_n(t)\phi_n(x)$ (Appendix \ref{Appendix2}). This gives the set of algebraic equations
	\begin{equation}\label{particle_dynamics}
	\dot{c}_n(t)=-\frac{in^2\pi^2\hbar}{2mL^2(t)}c_n(t)+\frac{\dot{L}(t)}{2L(t)}\bigg(c_n(t)+2\sum_{k}c_k(t)I_{nk}\bigg),
	\end{equation}
	where %$c_n(t)=\langle n|\psi(t)\rangle$ and
	\begin{equation}
	I_{nk}=\int x\,\phi_n^{*}(x)\,{\partial_x \phi_k(x)}\,dx.
	\end{equation}
%	with $\phi(x,t)=\sum_n c_n(t)\phi_n(x)$.	
	By combining Eqs.~\eqref{wall_Newton_Hamiltonian} and \eqref{particle_dynamics} we finally have a complete model for the system dynamics. In particular, by truncating the hierarchy of equations to the first $K$ eigenstates of $H(0)$ (corresponding to the smallest $K$ eigenvalues of such Hamiltonian), we obtain a system of $K$ differential equations that can be solved with standard numerical techniques.
	
	Eq.~\eqref{particle_dynamics} can also be written in the form of a von Neumann equation for the density matrix $\hat{\rho}$ of the particle
	\begin{equation}\label{vN}
	\frac{d\hat{\rho}}{dt}=-\frac{i}{\hbar}\left[\hat{H}^{*}(t), \hat{\rho}\right]
	\end{equation}
	with the effective Hamiltonian $\hat{H}^*(t)=\sum_n\hat{\cal H}_n(t)$, where we have introduced the operators
	\begin{equation}
	\hat{\cal H}_n(t)=\frac{n^2\pi^2\hbar}{2mL^2(t)}|n\rangle\langle n|+\frac{i\hbar\dot{L}(t)}{L(t)}\sum_{k\neq n}I_{nk}|n\rangle\langle k|.
	\end{equation}
	This helps us to model the effect of dissipative or dephasing mechanisms, the latter potentially being an important feature of the dynamics of a quantum particle interacting with a classical system. 

\begin{figure}[t!]
	\centering\includegraphics[width=\columnwidth]{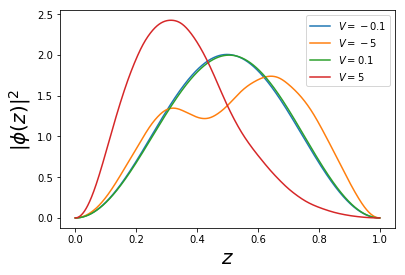}
	\caption{Particle wavefunction after a time $T={L_0}/{2V}$ when the wall is moving with constant speed $V$, for different values of $V$. The system dynamics is approximated considering only the first $20$ eigenstates of $\hat{H}(0)$.}\label{Figure:vconst}
\end{figure}

\section{Numerical simulations and discussion}\label{Numerical simulations}
We call $V=\dot{L}$ the speed of the wall and choose $L_0$ as the unit of length, $m$ as the unit of mass and $mL_0^2/\hbar$ as the unit of time. These units will be omitted in the following.

\begin{figure*}[t!]
	{\bf (a)}\hskip5cm{\bf (b)}\hskip5cm{\bf (c)}\\
	\includegraphics[width=0.66\columnwidth]{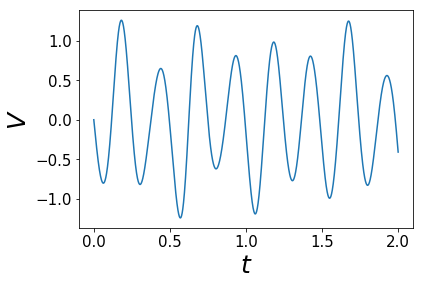}~\includegraphics[width=0.66\columnwidth]{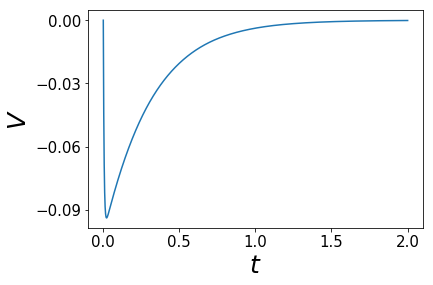}~\includegraphics[width=0.66\columnwidth]{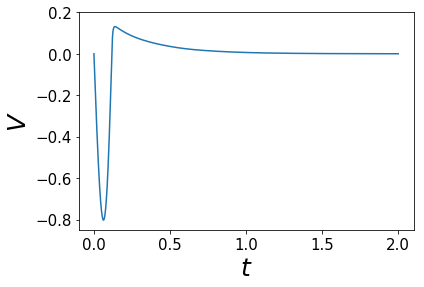}\\
	{\bf (d)}\hskip5cm{\bf (e)}\hskip5cm{\bf (f)}\\
	\includegraphics[width=0.66\columnwidth]{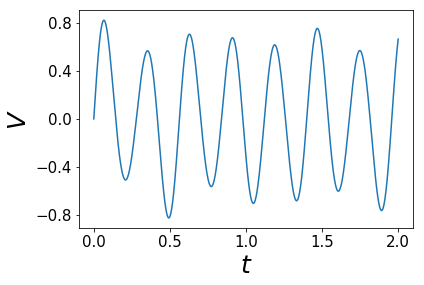}~\includegraphics[width=0.66\columnwidth]{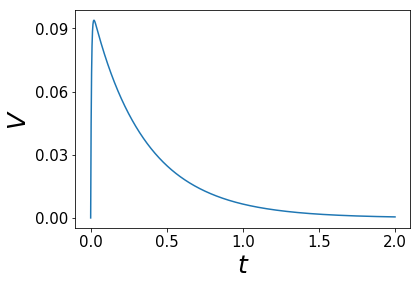}~\includegraphics[width=0.66\columnwidth]{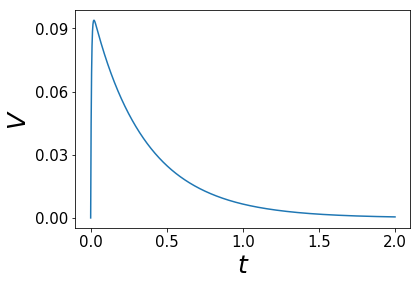}
	\caption{Wall speed as a function of time for initial compression and expansion when there is no friction [panel {\bf (a)} and {\bf (d)}, respectively]; under the presence of the frictional term $-\gamma V$ [panel {\bf (b)} and {\bf (e)}]; under friction described as in  Eq.~\ref{wall_Newton} [panel {\bf (c)} and {\bf (f)}]. The system's dynamics is approximated considering only the first $20$ eigenstates of $\hat{H}(0)$.}\label{Figure:friction}
\end{figure*}

We fix the mass of the wall to $M=0.05$. Although likely unrealistic, this choice helps us speeding up the evolution of the system and the numerical simulations, as it makes easier to accelerate the wall.

\subsection{Constant velocity}
Before simulating the dynamics of the combined system, we tested Eq.~\eqref{particle_dynamics} for the case of a uniform linear motion of the wall $L(t)=L_0+Vt$, when the particle is initially prepared in the ground state of its Hamiltonian. In Fig.~\ref{Figure:vconst} we show the unitarily transformed wavefunction $\phi(z,T)$ of the particle after a time $T={L_0}/{2V}$ for different values of $V$. It can be seen that the behaviour for both a slow compression and a slow expansion are in agreement with the adiabatic theorem, as the system is approximately found in the ground state of $\hat{H}(T)$. As expected, a fast expansion will causes the (not transformed) wavefunction $\psi$ to remain unchanged, while a unique behaviour can be seen in the case of a fast compression, where the system wavefunction is instead forced to compress with the box, as the accessible Hilbert space is reduced by the motion of the wall.

%\begin{figure*}[t!]
%	{\bf (a)}\hskip5cm{\bf (b)}\hskip5cm{\bf (c)}\\
%	\includegraphics[width=0.66\columnwidth]{Nofriction_Expansion}~\includegraphics[width=0.66\columnwidth]{2sidesFriction_Expansion}~\includegraphics[width=0.66\columnwidth]{halfFriction_Expansion}
%	\caption{Wall speed as a function of time for initial compression and expansion when there is no friction (Panel {\bf (a)} and Panel {\bf (b)} respectively), there is a $-\gamma V$ frictional force (Panel {\bf (c)} and Panel {\bf (d)}) and friction is present according to Equation~\ref{wall_Newton} (Panel {\bf (e)} and Panel {\bf (f)}). The system dynamics is approximated considering only the first $K=20$ eigenstates of $\hat{H}(0)$.}\label{Figure:frictionExpansion}
%\end{figure*}

\subsection{Effects of friction and decoherence}
Next, we study the effect of  friction on the motion of the wall. Plots of $V(t)$ are shown in Fig.~\ref{Figure:friction} for the case of a particle initially prepared in the ground state of its Hamiltonian. We consider $P_0/P(0)=1.1$ (initial compression) and $P_0/P(0)=0.9$ (initial expansion) when no friction is considered, when friction described as $-\gamma \dot L(t)$ is included in both the compression and expansion phases, and finally when friction affects the expansion phase through the term $-\gamma \dot{L}(t)\,h(\dot L)$, as for Eq.~\eqref{wall_Newton}. We take $\gamma=10$ (overdamped regime) and a total time of the evolution $T=2$. As expected, it can be seen how the presence of a classical friction is required in order to reach a stationary equilibrium point. 

Symmetric friction for both initial expansion and compression phase, and not symmetric one, i.e. present only in the initial expansion, makes the wall motion fairly simple to interpret: $V(t)$ never changes sign during the evolution, $|V(t)|$ grows due to the mismatch between the internal and the external pressure until it reaches a maximum, when it starts decreasing and asymptotically vanishes as the pressure difference becomes smaller and smaller while the frictional force keeps dumping the momentum of the wall.

A slightly more subtle behaviour is instead shown by the system when it is initially compressed and the friction acts only when the box is expanding. In such case, due to the absence of friction during the compression phase, the wall still carries momentum when the internal and external pressure compensate. Hence the system is still compressed after the pressure difference vanishes. When $V(t)$ reaches a minimum, the pressure difference has already changed sign, making the velocity grow to positive values (with the box expanding) until it reaches a maximum and then asymptotically vanishes due to the effect of the frictional force.

\begin{figure*}[t!]
	\hskip2cm{\bf (a)}\hskip8cm{\bf (b)}
	\includegraphics[width=\columnwidth]{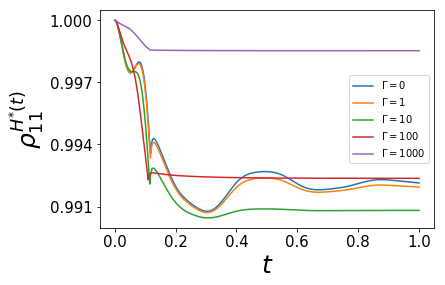}~\includegraphics[width=\columnwidth]{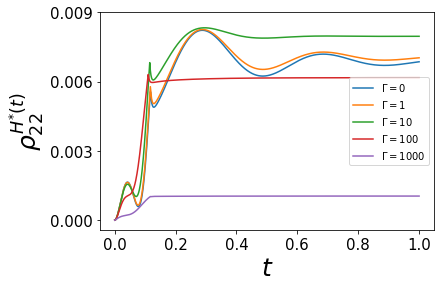}\\
	\hskip2cm{\bf (c)}\hskip8cm{\bf (d)}
%	{\bf (c)}\hskip3.5cm{\bf (d)}\\
	\includegraphics[width=\columnwidth]{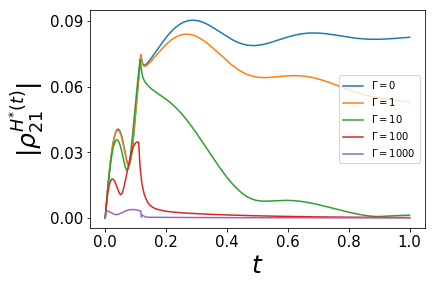}~\includegraphics[width=\columnwidth]{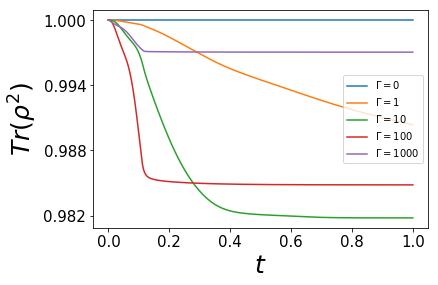}\\
	\caption{Panel {\bf (a)}: population of the ground state of $\hat{H}^*(t)$ as a function of time for different values of $\Gamma$. Panel {\bf (b)} [{\bf (c)}]: populations of the first excited state (coherences between the ground and of the first excited state) of $\hat{H^{*}}(t)$ as a function of time, for different values of $\Gamma$. Panel {\bf (d)}: purity of the state of the system against for various choices of $\Gamma$. The system dynamics is approximated considering only the first $40$ eigenstates of $\hat{H}(0)$.}\label{Figure:dephasing}
\end{figure*}
The latter renders the dynamics of the wall irreversible from a thermodynamic viewpoint. If one of two interacting quantum systems undergoes an irreversible dynamics, it is expected that both systems are subjected to some form of decoherence mechanism. However, due to the fact that we are assuming a classical dynamics for the wall, our model, alone, is incapable of capturing such phenomenon. Despite this limitation, we can still include a decoherence mechanism by introducing, phenomenologically, a non-unitary term $D\big[\hat{\rho}\big]$ in the effective dynamics entailed by Eq.~\eqref{vN} to obtain the Lindblad-like master equation
	\begin{equation}\label{deph}
	\frac{d\hat{\rho}}{dt}=-\frac{i}{\hbar}\big[\hat{H}^{*}(t), \hat{\rho}\big]+D\left[\hat{\rho}\right].
	\end{equation}
For the sake of simplicity we consider a pure dephasing -- at a rate $\Gamma$ -- in the instantaneous eigenbasis of the total Hamiltonian $\hat{H^{*}}(t)$ (Appendix \ref{Dephasing}).

In Fig.~\ref{Figure:dephasing} we show the effect of such mechanism on a system described by Eq.~\eqref{deph} and Eq.~\eqref{wall_Newton_Hamiltonian}, prepared in the ground state of its initial Hamiltonian, for different values of dephasing rate $\Gamma$ in the case of initial compression. A peculiar behavior is shown by the purity of the system state $\tr(\rho^2)$: contrary to what one would expect at first, increasing the value of $\Gamma$ does not always result into a further decrease of purity. This can be explained looking at the dynamics of the particle. The effect of $D\big[\hat{\rho}\big]$ is the destruction of the system coherences. In particular, the larger the value of $\Gamma$, the faster the coherences between the ground and the excited states, created by the unitary dynamics, are destroyed. This results in a smaller population transferred from the initial ground state to the excited states. Hence, if the value of $\Gamma$ is large enough,  the system is mostly kept in an eigenstate of the Hamiltonian (the ground state) by this particular dynamics, and less purity is lost at variance with the expectations of a faster loss of coherence induced by a strong dephasing.

\begin{figure}[h!]
	\includegraphics[width=0.9\columnwidth]{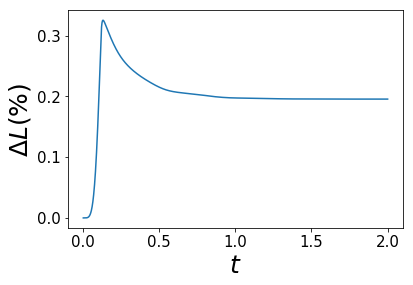}\\
	\caption{Relative difference in the length of the box with and without dephasing $\Delta L=[{L(\Gamma=10)-L(\Gamma=0)}]/{L(\Gamma=0)}$. The system dynamics is approximated considering only the first $40$ eigenstates of $\hat{H}(0)$.}\label{Figure:dephasingEq}
\end{figure}

Decoherence also affects the equilibrium length of the box, as shown in Fig. \ref{Figure:dephasingEq}. There, it can be seen that $L(t)$ for the system subjected to dephasing ($\Gamma=10$) is approaching a larger equilibrium value compared to the length of the box for the case of unitary dynamics. This shows how such mechanism can have observable macroscopic effects.

\subsection{Mechanical equilibrium and thermodynamics}
Finally we studied some thermodynamic properties of the system. We considered an initial thermal state with inverse temperature $\beta=0.1$. We assumed a unitary dynamics for the quantum system ($D\big[\hat{\rho}\big]=0$) and we set $\gamma=10$.

\begin{figure}[b!]
	{\bf (a)}\hskip3.5cm{\bf (b)}\\
	\includegraphics[width=0.5\columnwidth]{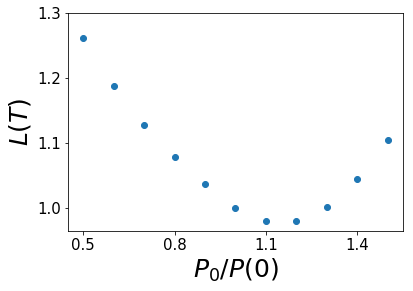}~\includegraphics[width=0.5\columnwidth]{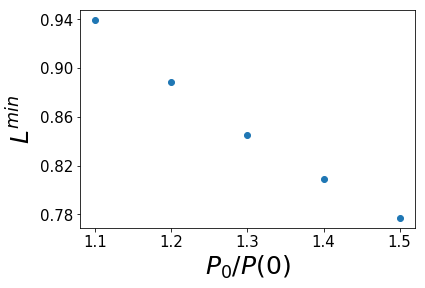}\\
	\caption{Panel {\bf (a)}: Final length of the box ($T=2$) as a function of $P_0/P(0)$. Panel {\bf (b)}: Minimum length of the box as a function of $P_0/P(0)$. The system dynamics is approximated considering only the first $20$ eigenstates of $\hat{H}(0)$.}\label{Figure:Leq}
\end{figure}

Fig.~\ref{Figure:Leq} shows the approximated equilibrium length of the box (length after a time $T=2$) in relation to the external pressure. Again, we can observe a peculiar behaviour when the system is initially compressed. The final length of the box can indeed be larger than the initial length. This is possible because of the absence of a frictional force in the compression phase combined with the fact that the external pressure does not depend on the position or the speed of the wall, contrary to the pressure due to the quantum particle. The evolution can be easily understood in two phases: the wall is compressed until a minimul length $L^{min}<L(0)$ is reached while the speed of the wall reaches zero and more energy is stored in the quantum system (work is done on the system); then a compression phase starts and both part of the initial internal energy and of the energy provided by the external gas in the compression phase contribute to the internal pressure making it possible for the wall to reach a final length $L(T)>L(0)$ (work is done by the system).

\begin{figure}[b!]
	{\bf (a)}\hskip3.5cm{\bf (b)}\\
	\includegraphics[width=0.5\columnwidth]{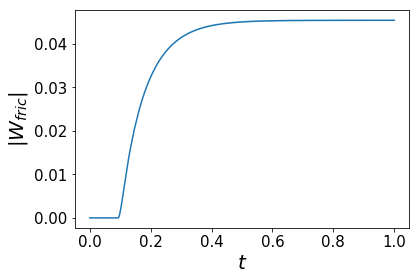}~\includegraphics[width=0.5\columnwidth]{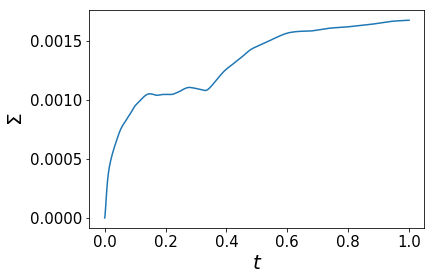}\\
	{\bf (c)}\hskip3.5cm{\bf (d)}\\
	\includegraphics[width=0.5\columnwidth]{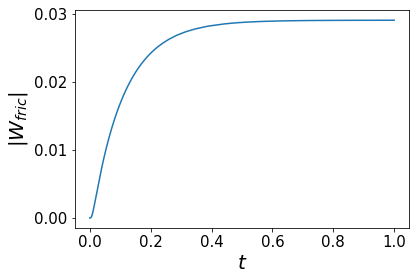}~\includegraphics[width=0.5\columnwidth]{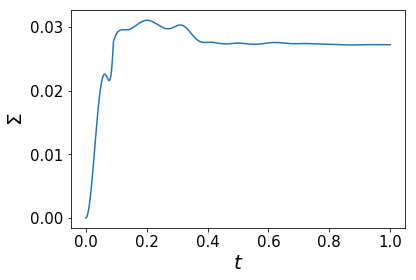}\\
	\caption{Energy dissipated due to the frictional force and quantum entropy production in the case of initial expansion [panels {\bf (a)} and {\bf (b)}] and compression [panels {\bf (c)} and {\bf (d)}]. The system dynamics is approximated considering only the first $20$ eigenstates of $\hat{H}(0)$.}\label{Figure:WS}
\end{figure}

In Fig.~\ref{Figure:WS} we show the entropy production of the quantum particle associated to the non-equilibrium process along with the total energy dissipated due to the classical friction~\cite{landipaternostro}
\begin{equation}
	|W_{fric}(t)|=\bigg\vert \int_{L(0)}^{L(t)}\gamma \dot{L} h(\dot{L}) dL\bigg\vert,
\end{equation}
for the cases of initial compression ($P_0/P(0)=1.1$) and initial expansion ($P_0/P(0)=0.9$).

The entropy production is calculated as the relative entropy \cite{Vedral2002} between the quantum state of the system and the instantaneous thermal equilibrium state $\rho^{eq}(t)=e^{-\beta \hat{H(t)}}/Z(t)$ with $Z(t)=\tr[e^{-\beta \hat{H(t)}}]$, partition function of the system \cite{DeffnerLutz, DeffnerLutzClosed, DeffnerCampbell}
\begin{equation}\label{rel_entr}
\Sigma(t)=S(\rho(t)||\rho^{eq}(t)),
\end{equation}
where $S(\sigma||\chi)=\tr[\sigma(\log\sigma-\log\chi)]$.

\begin{figure}[t!]
	{\bf (a)}\hskip3.5cm{\bf (b)}\\
	\includegraphics[width=0.5\columnwidth]{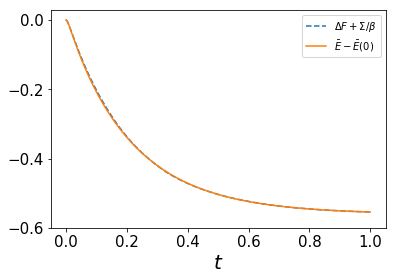}~\includegraphics[width=0.5\columnwidth]{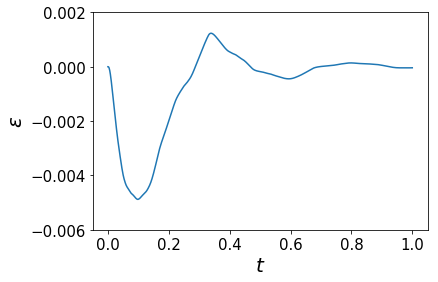}\\
	{\bf (c)}\hskip3.5cm{\bf (d)}\\
	\includegraphics[width=0.5\columnwidth]{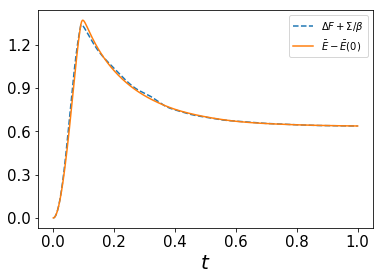}~\includegraphics[width=0.5\columnwidth]{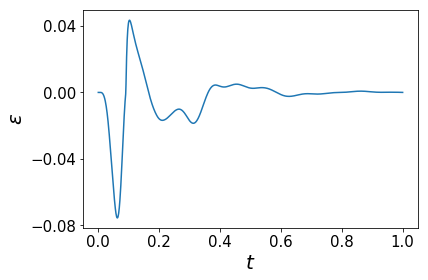}\\
	\caption{Change in the expectation value of the Hamiltonian of the particle and average energy change as expected from Eq.~\eqref{Wirr} [panels {\bf (a)} and  {\bf (c)}] and their difference [panels {\bf (b)} and {\bf (d)}] as functions of time in the case of initial expansion and compression. The system dynamics is approximated considering only the first $20$ eigenstates of $\hat{H}(0)$.}\label{Figure:irr}
\end{figure}

In Fig.~\ref{Figure:irr} we compare the energy change of the quantum system with the one expected if we define the irreversible work associated with the non-equilibrium process in terms of the relative entropy as $\Sigma(t)=\beta \langle W^{irr}(t) \rangle$ \cite{PhysRevLett.78.2690,DeffnerLutz,DeffnerLutzClosed}
\begin{equation}\label{Wirr}
\langle W^{irr}(t) \rangle=\langle W(t) \rangle - \Delta F(t),
\end{equation}
where $\langle W(t) \rangle$ is the average energy extracted from the quantum system and $\Delta F(t)$ is the Helmholtz free energy difference between the $\rho^{eq}(t)$ and $\rho(0)$.

Although the classical viscous friction is the only dissipative force in our model, the relationship between its work and the entropy production of the quantum system is non-trivial. Indeed, these two quantity are, in principle, a measure of two different concepts of irreversible work. While the work of the frictional force corresponds to the energy physically dissipated into an environment, the irreversible work associated to the relative entropy describes the wasted energy that is not being extracted from the system and that could have been extracted instead with an equilibrium, quasi-static, transformation. The last, however, is not actually dissipated and, if the system is kept isolated, more energy can still be retrieved by means of another unitary transformation. The same conclusion can be drown by observing that the final state of the system is non-passive \cite{NatCom2018, Allahverdyan2004}. Notice, however, that, althought different, both the relative entropy and the work done by the viscous frictional force are expected to depend on the speed of the transformation.

\begin{figure}[t!]
	{\bf (a)}\hskip3.5cm{\bf (b)}\\
	\includegraphics[width=0.5\columnwidth]{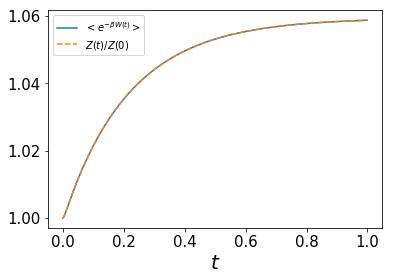}~\includegraphics[width=0.5\columnwidth]{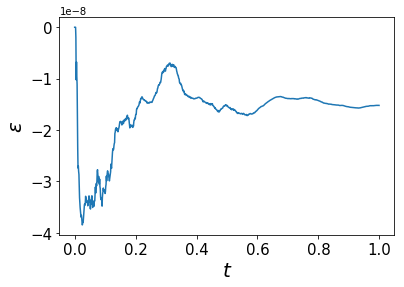}\\
	{\bf (c)}\hskip3.5cm{\bf (d)}\\
	\includegraphics[width=0.5\columnwidth]{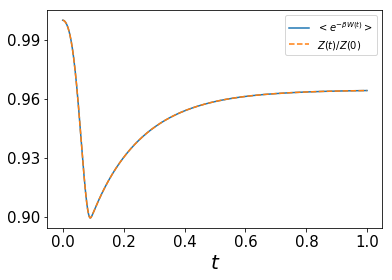}~\includegraphics[width=0.5\columnwidth]{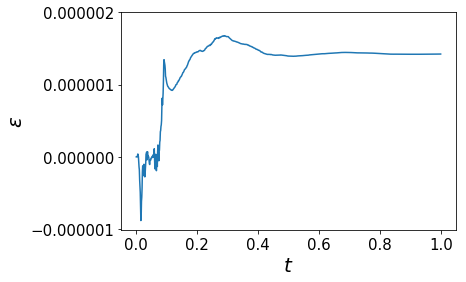}\\
	\caption{Left- and right-hand sides of Eq.~\eqref{Jarzynski} [panels {\bf (a)} and {\bf (c)}] and their difference [panels {\bf (b)} and {\bf (d)}] as functions of time in the case of initial expansion and compression. The system's dynamics is approximated considering only the first $20$ eigenstates of $\hat{H}(0)$.}\label{Figure:Jarz}
\end{figure}

Finally, we verified numerically (Appendix \ref{Numerical verification of the quantum Jarzynski equality}) the validity of the Jarzynski equality \cite{PhysRevLett.78.2690, QuantumWork, DeffnerCampbell, VinjanampathyAnders}
\begin{equation}\label{Jarzynski}
	\langle e^{-\beta W(t)} \rangle=e^{-\beta \Delta F(t)}
\end{equation}
for the quantum system, as shown in Fig.~\ref{Figure:Jarz}.
The length of the box and the speed of the wall depend on the interactions between the quantum and the classical system and are unknown before the evolution. However, their time dependence can be observed during the dynamics, and, once known, they can be treated as time dependent parameters of the Hamiltonian and the standard formulation of the Jarzynski equality holds.

\section{Considerations an limitations}\label{Considerations an limitations}
The main assumption of our model is the premise that the length of the box $L(t)$ can be considered a classical variable, undergoing classical dynamics. Current description of thermodynamic processes involving work extraction from a closed quantum working medium rely on the assumption that such processes can be described making use of time dependent parameters of the Hamiltonian. Our approach shares this premise but considers the back-action of the quantum system to determine these classical parameters in order to describe reciprocal interactions between the quantum and the classical system and model a spontaneous, non-driven, transformation. This back-action is introduced via the radiation pressure, which depends on an observable of the quantum system. Overcoming this limitation would involve a full-microscopic description of the intercting systems and would require to address the challenging task of redefining thermodynamic concepts such as heat and work. A first step to address this problem could be the introduction of fluctuations on the parameter $L(t)$. However, the inclusion of the dynamics of such fluctuations is not necessarily a straightforward extension of the model and could be a complex task on its own.

The absence of friction when the box length is decreasing causes multiple unexpected behaviours of the system. Adding friction in the compression phase would probably make the model more realistic. To do so, one should know both the expression of the frictional force and the effects of such mechanism on the quantum system. The latter cannot be ignored, as the presence of a frictional force is usually associated with heat exchange. Depending on the specific mechanism, both effects could be included in the model by adding the force in Eq.~\eqref{wall_Newton} and the correct dissipator $D[\hat{\rho}]$ in Eq.~\eqref{deph}.

\begin{figure}[t!]
	{\bf (a)}\\
	\includegraphics[width=0.8\columnwidth]{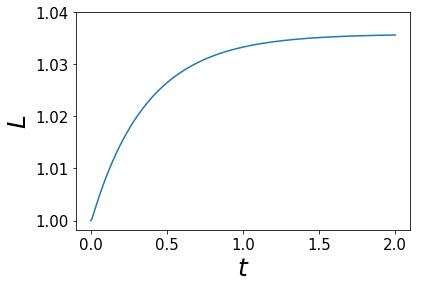}\\
	{\bf (b)}\\
	\includegraphics[width=0.8\columnwidth]{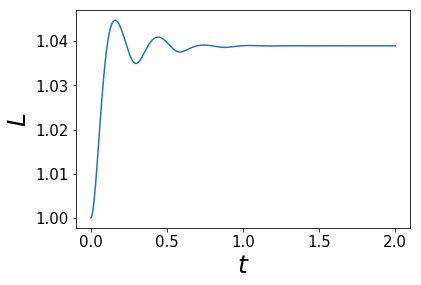}\\
	{\bf (c)}\\
	\includegraphics[width=0.8\columnwidth]{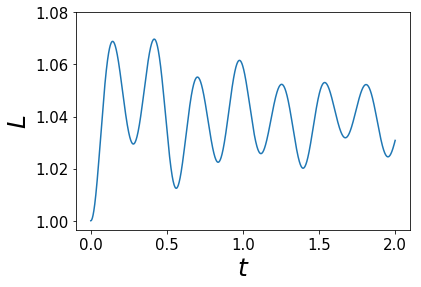}\\
	\caption{Length of the box in the case of initial expansion ($P_0/P(0)=0.9$) as a function of time when the system is prepared in the ground state of its Hamiltonian, for $\gamma=10$ [panel {\bf (a)}], $\gamma=1$ [panel {\bf (b)}] and $\gamma=0.1$ [panel {\bf (c)}]. The system dynamics is approximated considering only the first $20$ eigenstates of $\hat{H}(0)$.}\label{Figure:osc}
\end{figure}

\begin{figure*}[t!]
	{\bf (a)}\hskip3.5cm{\bf (b)}\hskip3.5cm{\bf (c)}\hskip3.5cm{\bf (d)}\\
	\includegraphics[width=0.52\columnwidth]{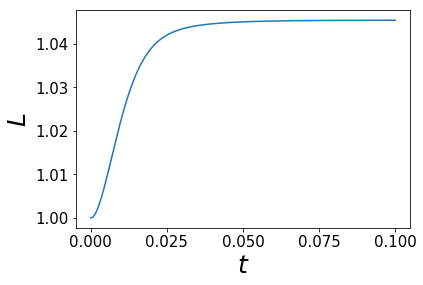}~\includegraphics[width=0.52\columnwidth]{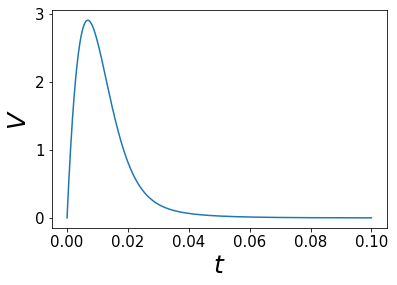}~\includegraphics[width=0.52\columnwidth]{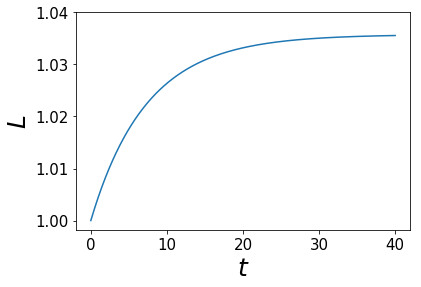}~\includegraphics[width=0.52\columnwidth]{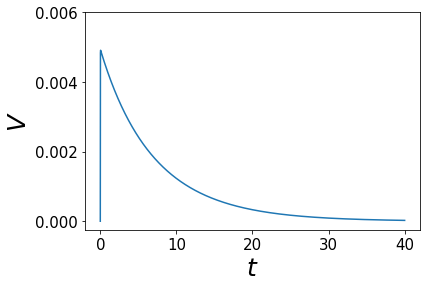}\\
	\caption{Length of the box and speed of the wall when $M=0.001$ [panels {\bf (a)} and {\bf (b)}] and $M=1$ [panels {\bf (c)} and {\bf (d)}] as functions of time in the case of initial expansion ($P_0/P(0)=0.9$). The system dynamics is approximated considering only the first $20$ eigenstates of $\hat{H}(0)$.}\label{Figure:Mass}
\end{figure*}

Regarding the choice of the parameters, both $\gamma$ and $M$ affects how the system reaches the equilibrium point. As shown in Fig.~\ref{Figure:osc}, decreasing the value of $\gamma$ leads to oscillations in the length of the box. In Fig.~\ref{Figure:Mass} it can be seen how increasing the mass of the wall from $M=0.001$ to $M=1$ while staying in the overdamped regime ($\gamma=10 M/0.05$) increases the time required for the system to reach mechanical equilibrium. It is also worth noticing that, as increasing $M$ slows down the dynamics, it also makes the evolution closer to a quasi-static transformation, with important thermodynamic consequences. To see the effect on the adiabaticity of the transformation, we prepared the system in the first eigenstate of the initial Hamiltonian in the case of initial compression ($P_0/P(0)=0.9$) and we calculated the fidelity of the state of the system at each time step with the corresponding ground-state of the time-dependent Hamiltonian of the system. The minimum fidelity reached durig the evolution is $\approx0.997$ for $M=0.001$ while it is still $\approx 1.000$, within numerical error, for $M=1$.

As a last note, we would like to point out that our model describes a work extraction protocol from a quantum working medium without the need to introduce any energy measurement on the system. Despite this, the Jarzynski equality derived from a two measurement approach \cite{DeffnerCampbell, QuantumWork, RevModPhys.83.771} still holds. However, the idea of realizing such transformation without carrying out any measurement on the quantum system has to be taken carefully. In our model, the static mechanical equilibrium point is reached because of the presence of friction in the classical dynamics. Due to entanglement between the wall and the particle, in a full quantum version of the evolution, such dissipative mechanism would involve some form of decoherence in both systems. This would imply that, in order to make the model more realistic, the associated dissipator has to be included in Eq.~\eqref{deph}, as discussed in Sec.~\ref{Numerical simulations}, and hence it is possible that such mechanism would play the role of some unread measurements on the system, in the spirit of a quantum dynamical model of measurement (cf. Ref.~\cite{Allahverdyan} for a review). 

\section{Conclusions}
We have presented a case-study of a closed quantum system undergoing a spontaneous thermodynamic transformation. In particular, we have studied a simple model of a quantum particle inside a piston with an insulating wall interacting with a classical gas on the outside. A mismatch between internal and external pressure on the piston causes the two gasses to be out of mechanical equilibrium and induces the thermodynamic transformation.

We have shown how the presence of a frictional force is required in order to reach a static mechanical equilibrium point and how decoherence can affect the system dynamics with macroscopic consequences. We have studied how the mechanical equilibrium point is reached as a function of the initial internal and external pressure. We have looked at the irreversibility of the process, separating the physical energy dissipation due to the presence of the classical friction from the entropy production of the closed quantum system, related to the potential work wasted by a non-equilibrium transformation. We have verified the validity of the Jarzynski equality for our quantum system.

Our approach allows us to describe spontaneous thermodynamic transformations of a closed quantum system without a-priori knowledge of the time-dependence of its Hamiltonian, as long as the quantum system is interacting with a classical system and a model of the interaction is known. Our model is capable of capturing various instances of the thermodynamic transformation considered and no direct measurement on the quantum system is required in order to extract energy from it in the form of work.

However, certain limitations are still present, such as the phenomenological inclusion of frictional forces in the classical dynamics and decoherence mechanisms in the quantum system, a procedure that ignores any inter-relationship between such mechanisms. To overcome such limitations, a full quantum description of the thermodynamic process is required. This would also help us understand to what extent work extraction protocols can be described in terms of changes in time dependent parameters of the system Hamiltonian.

A meaningful step to extend our model in the direction of a completely quantum description would be the inclusion of fluctuations in the classical parameter $L(t)$.

\acknowledgments

We acknowledge the support by the SFI-DfE Investigator Programme (grant 15/IA/2864), the European Union's Horizon 2020 FET-Open project  TEQ (766900), the Leverhulme Trust Research Project Grant UltraQuTe (grant RGP-2018-266), the Royal Society Wolfson Fellowship (RSWF/R3/183013), and the UK EPSRC (EP/T028424/1). 

%\newpage
%\bibliographystyle{apsrev4-1.bst}
	\bibliography{bibl} 
%	\bibliographystyle{ieeetr}

%\newpage
%\
%\newpage
\appendix
\section{Quantum pressure}\label{Appendix1}
The pressure exerted by a classical ideal gas of particles of mass $m$ in a box of length $L$ and section $\Sigma$ is easily calculated:  the mean impulsive force applied on the wall by a particle is given by the ratio between the momentum transferred in a particle-wall collision and the time between two consecutive collisions. If we restrict the motion of the particles to one dimension, we have
\begin{equation}
P^{cl}=\frac{\langle p^2\rangle}{mL\Sigma},
\end{equation}
where $\langle p^2\rangle$ is the mean square momentum of the particles.

A similar expression also holds if we consider $P$ to be the pressure exerted by a wavepacket corresponding to the wavefunction of the particle, analogous to the well-known radiation pressure. If the plane-wave component of the packet with wavenumber $k$ has amplitude $|c_k|^2$, intensity $I_k$, speed $v_k$ and we assume the wall to be a perfectly reflecting surface, we get
\begin{equation}
P^{rad}=\sum_{k}|c_k|^2\frac{2I_k}{v_k}=2\sum_{k}|c_k|^2u_k,
\end{equation}
where
\begin{equation}
u_k=\frac{\hbar^2k^2}{2mL\Sigma}=\frac{p_k^2}{2mL\Sigma},
\end{equation}
is the energy density of the corresponding wavenumber $k$. Hence we have
\begin{equation}
P^{rad}=\frac{\langle p^2\rangle}{mL\Sigma}.
\end{equation}
These expressions suggest us to assume a formally similar definition for the pressure exerted by the quantum particle on the wall
\begin{equation}
P=\frac{\langle \hat{p}^2\rangle}{mL\Sigma},
\end{equation}
where now $\langle \hat{p}^2\rangle$ is the expectation value of the observable $\hat{p}^2$.

\section{Equation of motion in the eigenbasis of the initial Hamiltonian}\label{Appendix2}
Writing Eq.~\eqref{fixed_bound} in terms of the state vector $|\phi\rangle$ such that $\phi(x)=\langle x\ket{\phi}$, and projecting both sides of the equation upon $\langle n|$ we obtain
\begin{equation}
\langle n|i\hbar\frac{d}{dt}|\phi\rangle=\frac{1}{L^2}\frac{\langle n|\hat{p}^2|\phi\rangle}{2m}-\frac{\dot{L}}{2L}\langle n|(\hat{x}\hat{p}+\hat{p}\hat{x})|\phi\rangle,
\end{equation}
where for instance
\begin{equation}
\langle n|\hat{x}\hat{p}|\phi\rangle=-i\hbar\int x\, \phi_n^{*}(x)\frac{\partial \phi(x,t)}{\partial x}dx.
\end{equation}
Expanding $\phi(x,t)$ in the eigenbasis of $H(0)$, we obtain
\begin{equation}
\langle n|\hat{x}\hat{p}|\phi\rangle=-i\hbar\sum_k c_k(t) \int x\, \phi_n^{*}(x)\,\frac{\partial \phi_k(x)}{\partial x}\,dx,
\end{equation}
where $\phi(x,t)=\sum_k c_k(t)\phi_k(x)$. Similarly, we get
\begin{equation}
\langle n|\hat{p}\hat{x}|\phi\rangle=-i\hbar c_n(t) -i\hbar\sum_k c_k(t)\int x\,\phi_n^{*}(x)\,\frac{\partial \phi_k(x)}{\partial x}\,dx.
\end{equation}
As the eigenstates of $H(0)$ have wavefunctions $\phi_k(x)=\sqrt{2}\sin(k\pi x)$, the elements of the matrix
\begin{equation}
I_{nk}=\int x\, \phi_n^{*}(x)\frac{\partial \phi_k(x)}{\partial x}\, dx
\end{equation}
have analytical expressions that can be easily calculated before solving the dynamics. This makes the dynamics easier to be solved numerically compared to a mere application of Eq.~\eqref{fixed_bound}.

\section{Dephasing}\label{Dephasing}
We want to include pure dephasing in the basis of $\hat{H^*}(t)$. In the eigenbasis of such operator, we consider
\begin{equation}
D\big[\hat{\rho}\big]_{n(t)k(t)}=-\frac{\Gamma}{2}\rho_{n(t)k(t)}(1-\delta_{n(t)k(t)}).
\end{equation}
In order to implement this mechanism in our numerical simulations while still working on the eigenbasis of the initial Hamiltonian, at each time steps we first find the eigenstates of the total Hamiltonian of the system and transform $\hat{\rho}(t)$ in such basis, obtaining the matrix $\hat{\rho^{H^{*}(t)}}(t)$. Then, we keep only the non-diagonal elements of $\hat{\rho^{H^{*}(t)}}(t)$ to build $D\big[\hat{\rho^{H^{*}(t)}}(t)\big]_{n(t)k(t)}$ and transform this back to the eigenbasis of the initial Hamiltonian, obtaining $D\big[\hat{\rho}(t)\big]_{nk}$.

\section{Numerical verification of the quantum Jarzynski equality}\label{Numerical verification of the quantum Jarzynski equality}
In order to verify the Jarzynski equality [Eq.~\eqref{Jarzynski}], we have to calculate the ensamble average
\begin{equation}\label{eW}
	\langle e^{-\beta W}(t) \rangle=\int dwP(w,t)e^{-\beta w},
\end{equation}
where $P(w)$ is the probability density of the work distribution \cite{QuantumWork}
\begin{equation}
P(w,t)=\sum_{n,m}\delta(w-(E_m(t)-E_n(0))p(m,t|n)p_n,
\end{equation}
with $p_n$ the probability of measuring the $n^{th}$ eigenstate of the initial Hamiltonian at the start of the evolution, $E_n(0)$ the associated eigenvalue, $E_m(t)$ the $m^{th}$ eigenvalue of the system Hamiltonian at time $t$, and $p(m,t|n)$ the probability of measuring such value of the energy if the system is found in the $n^{th}$ eigenstate of the initial Hamiltonian at the start of the evolution.

\begin{figure}[h!]
	\centering
	{\bf (a)}\\
	\includegraphics[width=\columnwidth]{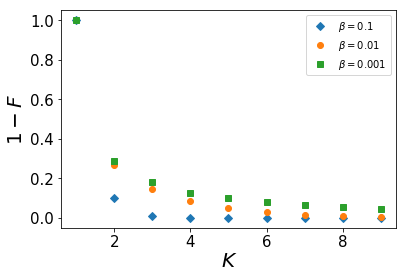}\\{\bf (b)}\\ \includegraphics[width=\columnwidth]{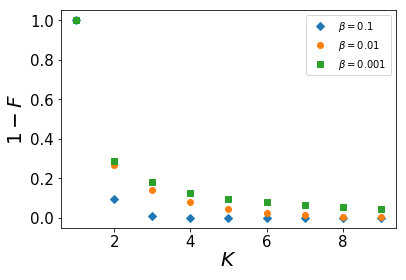}\\
	\caption{Infidelity between the final state reached by truncating to the first $K$ eigenstates of $H(0)$ and to the final state reached by truncating to the first $K+1$ eigenstates of $H(0)$ for different values of initial inverted temperature $\beta$ in the case of initial compression [panel {\bf (a)}] and expansion [panel {\bf (b)}].}\label{convergence}
\end{figure}

In order to calculate the integral in Eq.~\eqref{eW}, we have to evaluate $p(m,t|n)$. To do so, we first simulate the system dynamics according to Eq.~\eqref{wall_Newton_Hamiltonian} and Eq.~\eqref{particle_dynamics}. Then, once $L(t)$ and $V(t)$ are known, we can discard the dynamics of the classical system and consider only the quantum system. We hence solve the dynamics of the quantum system for each eigenstate $|n\rangle$ of the initial Hamiltonian and calculate the correspondings conditional probabilities $p(m,t|n)$ for each eigenstate $|m(t)\rangle$ of the system time-dependent Hamiltonian.
\section{Truncation of the Hilbert space and conergence}\label{Truncation of the Hilbert space and conergence}
In order to solve the system dynamics numerically, we truncate Eq.~\eqref{particle_dynamics} to the first K eigenstates of $H(0)$. In our simulations we used $K\geq20$. Here we show that this approximation is reliable for our system by looking at the convergence of the solution depending on $K$. In paricular, we consider the \textit{Infidelity} $1-F$ between the final states of the system (after a time $T=2$) obtained by considering only the first $K$ eigenstates of $H(0)$, and the first $K+1$ eigenstates of $H(0)$ (where the \textit{Fidelity} $F$ between two quantum states $|\psi\rangle$, $|\phi\rangle$ is definded as $F=|\langle\psi|\phi\rangle|^2$).
In Figure \ref{convergence} we show how fast the solution converges for different initial temperatures, which was found to be the parameter with the most significant influence on the rate of convergence.

%\onecolumngrid

\end{document}